\documentclass[12pt]{iopart}

\usepackage{iopams}

\newcounter{defin}
\newcounter{lemma}
\newcounter{theorem}
\newcounter{proposition}
\newcounter{example}
\newenvironment{lemma}{\par\refstepcounter{lemma}     \textbf{Lemma \thelemma.} }{\rm\par}
\newenvironment{theorem}{\par\refstepcounter{theorem}     \textbf{Theorem \thetheorem.}\ }{\rm\par}
\newenvironment{proposition}{\par\refstepcounter{proposition}     \textbf{Proposition \theproposition.}\ }{\rm\par}
\newenvironment{definition}{\par\refstepcounter{defin}     \textbf{Definition \thedefin.}\ }{\rm\par}

\begin{document}

\title{Information capacity of continuous variable measurement channel}
\author{A S Holevo and  A A Kuznetsova}

\address{Steklov Mathematical Institute, Gubkina 8, 119991 Moscow, Russian Federation}
\eads{\mailto{holevo@mi-ras.ru}, \mailto{kuznetsova.a.a@bk.ru}}

\begin{abstract}
The present paper is devoted to investigation of the classical capacity of
infinite-dimensional continuous variable quantum measurement channels. A number of usable conditions are introduced
that enable us to apply previously obtained general results to specific
models, in particular, to the multi-mode bosonic Gaussian measurement
channels. An explicit formula for the classical capacity of the Gaussian
measurement channel is obtained in this paper without assuming the global
gauge symmetry, solely under certain ``threshold condition''. The
result is illustrated by the capacity computation for one-mode
squeezed-noise heterodyne measurement channel.
\end{abstract}



\maketitle

\section{Introduction}

From the viewpoint of information theory measurements are peculiar communication
channels that transform input quantum states into classical output data. As such,
they are described by the information capacity which is the most important
quantity characterizing their ultimate information-processing performance.
The present work develops investigation of the capacities of quantum
measurement channels, initiated in ~\cite{hall,da,Oreshkov,h5}. The emphasis here is
on the measurements with continuous multi-dimensional output. It is well known
(see, e.g., \cite{h5}) that channels with continuous classical output (in
contrast to discrete output) in principle cannot be
extended to quantum channels (i.e. maps with the quantum input and output).
This prevents from a direct use of the well developed quantum Shannon theory
for channels and thus requires a separate study. In particular, this
remark fully applies to bosonic Gaussian measurement channels, which are
of great both theoretical and practical interest.

In \cite{Hybrid-2} a proof of the coding theorem for the classical capacity
of a measurement channel with arbitrary output alphabet was given in the
most general setting. In the present work a number of applicable conditions
are developed that make it possible to implement the general results, in
particular, to bring the capacity calculations to the quite specific
formulas. In proposition \ref{p1} of section \ref{s1} a convenient
expression is obtained for the energy-constrained classical capacity of a
measurement channel in terms of the differential entropy. This expression is
further specified in proposition \ref{p2} for irreducibly covariant measurement
channels. Section \ref{s2} is devoted to Gaussian measurement channels.
Theorem \ref{tt2} is proved, giving an explicit formula for the
energy-constrained classical capacity of the Gaussian measurement channel
without assuming the global gauge symmetry, solely under the
\textquotedblleft threshold condition\textquotedblright\ (\ref{thr}). This
result generalizes theorem 1 from \cite{acc}, in which the globally
gauge-covariant case was considered. The result is illustrated by an example
of the single-mode squeezed-noise heterodyne measurement.

\section{Capacity of the measurement channel and the output differential entropy}

\label{s1}

Let $\mathcal{H}$ be a separable Hilbert space. By $\mathfrak{B}(\mathcal{H}
) $ we denote the algebra of all bounded operators in $\mathcal{H}$, $
\mathfrak{T}(\mathcal{H})$ is the Banach space of trace-class operators, $
\mathfrak{S}(\mathcal{H})$ is the convex subset of \textit{density operators},
 i.e. positive operators with unit trace, also called \textit{quantum states}.

We also introduce a measure space $(\Omega ,\mathcal{F},\mu ),$ where $
\Omega $ is a complete separable metric space, $\mathcal{F}$ is a $\sigma $
-algebra of its subsets, $\mu $ is a $\sigma $-finite measure on $\mathcal{F}
$.

\begin{definition}
\label{POVM} \textit{Probability operator-valued measure (POVM)} on $\Omega$
is a family $M=\{ M(A), A \in \mathcal{F}\}$ of bounded Hermitian operators
in $\mathcal{H}$, satisfying the following conditions:

$1.$ $M(A) \geq 0; A \in \mathcal{F}$;

$2.$ $M(\Omega)=I$, where $I$ is the unit operator in $\mathcal{H}$;

$3.$ for arbitrary countable decomposition $A=\cup A_{i},(A_{i}\in \mathcal{F
},\,A_{i}\cap A_{j}=\emptyset ,i\neq j)$ the relation $M(A)=\sum_{i}M(A_{i})$
holds in the sense of weak convergence of operators.

The POVM defines a \textit{quantum observable} with values in $\Omega$.
\textit{The probability distribution} of the observable $M$ in the state $
\rho$ is given by the formula
\begin{equation}  \label{distrib}
P_\rho(A)=\mathrm{Tr} \rho M(A),\quad A \in \mathcal{F}.
\end{equation}\end{definition}
For brevity, we sometimes write $P_\rho(d\omega)=\mathrm{Tr} \rho M(d\omega)$.
As it is known, there is a unique POVM, for which $P_\rho$ is given
by the formula (\ref{distrib}).

\begin{definition}
\label{M-ch} \textit{Measurement channel} $\mathcal{M}$
is an affine map $\rho\rightarrow P_\rho(d\omega)$ of the convex set of quantum states $
\mathfrak{S}(\mathcal{H})$ into the set of probability distributions on $
\Omega$.
\end{definition}

The purpose of this work is to study the information characteristics of the
channel $\mathcal{M}$, in particular, its capacity for transmitting the
classical information, with a natural energy restriction at the input,
basing on general expressions obtained previously in \cite{h1}, \cite
{Hybrid-2}.

\begin{lemma}
\label{density} \cite{Hybrid-2} \textit{For arbitrary observable $M(d\omega
) $ with values in $\Omega $ there exists a $\sigma $-finite measure $\mu $
on $\Omega $, such that for any density operator $\rho \in \mathfrak{S}(
\mathcal{H})$ the probability distribution $P_{\rho }(d\omega )=\mathrm{Tr}
\rho M(d\omega )$ has a density $p_{\rho }(\omega )$ with respect to measure
$\mu $.}
\end{lemma}

Therefore the measurement channel can be considered as an affine map $
\mathcal{M}:\rho \rightarrow p_{\rho }(\omega ),$ and we will write $p_{\rho
}=\mathcal{M}(\rho ).$

\begin{definition}
\label{ensemble} A Borel probability measure $\pi $ on $\mathfrak{S}(
\mathcal{H})$ will be called \textit{ensemble}. The \textit{average state}
of the ensemble $\pi $, determined by the Bochner integral in $\mathfrak{T}(
\mathcal{H})$ \cite{ContEns}
\begin{equation}
\bar{\rho}_{\pi }=\int\limits_{\mathfrak{S}(\mathcal{H})}\rho \pi (d\rho ),
\label{average}
\end{equation}
is just the barycenter of the measure $\pi .$
\end{definition}

Note that in specific applications the measure $\pi$ is usually concentrated
on some parametrically given subset of states. In these cases, it is more
convenient to define the ensemble as a pair $\{\pi (dx), \rho_x\}$, where
the parameter $x$ runs over a complete separable metric space $X$, $\pi(dx)$
is a probability measure on $X$, $\{ \rho_x, x\in X\}$ is a measurable
family of states.

For a given ensemble $\pi $ and a measurement channel $\mathcal{M}$ the
Shannon mutual information between the input and the output of the channel
can be defined by the formula
\begin{equation*}
I(\pi ,\mathcal{M})=\int_{\mathfrak{S}(\mathcal{H})}h(\mathcal{M}(\rho
)\Vert \mathcal{M}(\bar{\rho}_{\pi }))\pi (d\rho ),
\end{equation*}
where
\begin{equation*}
h(\mathcal{M}(\rho )\Vert \mathcal{M}(\bar{\rho}_{\pi }))=\int_{\Omega
}p_{\rho }(\omega )\log \frac{p_{\rho }(\omega )}{p_{\bar{\rho}_{\pi
}}(\omega )}\mu (d\omega )
\end{equation*}
is the classical relative entropy. The functional $I(\pi ,
\mathcal{M})$ is well defined and takes values in $[0;+\infty ].$

We introduce the generalized differential entropy of a probability density $p(\omega )$
on $(\Omega ,\mu )$ by the relation  (cf. \cite{cover}, where $\mu$ is the Lebesgue measure on $\mathbb{R}^{s}$)
\begin{equation*}
h(p)=-\int_{\Omega }p(\omega )\log p(\omega )\mu (d\omega ),
\end{equation*}
provided the integral exists, including the values $\pm\infty$. If $|h(
\mathcal{M}(\bar{\rho}_{\pi }))|<\infty $ then
\begin{equation*}
I(\pi ,\mathcal{M})=\int_{\mathfrak{S}(\mathcal{H})}\int_{\Omega }
p_{\rho }(\omega )\left(\log {p_{\rho }(\omega )}-\log {p_{\bar{\rho}_{\pi
}}(\omega )}\right) \mu (d\omega )\pi (d\rho )
\end{equation*}
\begin{equation*}
=h(\mathcal{M}(\bar{\rho}_{\pi }))-\int_{\mathfrak{S}(\mathcal{H})}h(
\mathcal{M}(\rho ))\pi (d\rho ),
\end{equation*}
where $h(\mathcal{M}(\rho ))=h(p_{\rho })$ is the differential entropy of
the channel $\mathcal{M}$ output probability density.

In the case of infinite-dimensional $\mathcal{H}$ one usually introduces a
constraint onto the input states of the channel (otherwise the capacity may
be infinite). Let $H$ be a positive self-adjoint (in general unbounded)
operator in the space $\mathcal{H}$, with the spectral decomposition $
H=\int_{0}^{\infty }\lambda dP(\lambda ),$ where $P(\lambda )$ is the
spectral function. In specific applications, the role of $H$ is played by
the energy operator of a quantum system at the input of the channel. We
introduce the subset of states
\begin{equation}
\mathfrak{S}_{E}=\{\rho :~\mathrm{Tr}\rho H\leq E\},  \label{setE}
\end{equation}
where $E$ is a positive constant, and the trace is understood as the
integral (for more detail see \cite{ClCap})
\begin{equation*}
\mathrm{Tr}\rho H=\int_{0}^{\infty }\lambda d(\mathrm{Tr}\rho P(\lambda )).
\end{equation*}

Consider the energy constraint on the input ensemble $\pi $ defined by the
relation $\mathrm{Tr}\bar{\rho}_{\pi }H\leq E$. According to theorem 1 from
\cite{Hybrid-2}, the classical capacity of the measurement channel $\mathcal{
M}$ with the input constraint is given by the relation
\begin{equation*}
C(\mathcal{M},H,E)=\sup\limits_{\pi ^{\prime }:~\mathrm{Tr}\bar{\rho}_{\pi
^{\prime }}H\leq E}I(\pi ^{\prime },{\mathcal{M}}),
\end{equation*}
where $\pi ^{\prime }=\{{\pi ^{\prime }}_{x};\rho _{x}^{\prime }\}$ runs
through the finite ensembles. Then, obviously,
\begin{equation}  \label{lek}
C(\mathcal{M},H,E)\leq \sup_{\pi :~\mathrm{Tr}\bar{\rho}_{\pi }H\leq E}I(\pi
,\mathcal{M}),
\end{equation}
where the supremum is taken over all ensembles.

\begin{proposition}
\label{p1} \textit{Let $\mathcal{M}:~ \rho \rightarrow p_\rho(\omega)$ be a
measurement channel such that for any $\rho$ the density $p_\rho(\omega)$ is
bounded. \newline
\indent Let a nonnegative function $H_c(\omega)$ be given on $\Omega$ that satisfies
the conditions
\begin{equation}  \label{cond1}
\int_{\Omega} e^{-\theta H_c(\omega)} \mu(d\omega) <\infty,
\end{equation}
for some $\theta>0$, and
\begin{equation}  \label{cond2-}
\int_{\Omega} H_c(\omega) p_\rho (\omega) \mu(d\omega) <\infty, \quad\forall
\rho \in \mathfrak{S}_E,
\end{equation}
then
\begin{equation}  \label{entfin-}
h(p_\rho) < \infty,\quad \forall \rho \in \mathfrak{S}_E.
\end{equation}
\indent In this case, the capacity $C(\mathcal{M}, H, E) $ is given by the formula
\begin{equation}  \label{cmfe}
C(\mathcal{M}, H, E) = \sup_{\pi: ~ \mathrm{Tr} \bar{\rho}_{\pi} H \leq E}
\left[h(\mathcal{M}(\bar{\rho}_{\pi})) - \int_{\mathfrak{S}(\mathcal{H})} h(
\mathcal{M}(\rho))\pi(d\rho)\right],
\end{equation}
where the supremum is taken over all ensembles.}
\end{proposition}

\textit{Proof.} It is shown in \cite{acc} that the value $-\infty$ is excluded
for the output differential entropy due to the fact that the density $p_\rho$
is bounded. We show that under the condition (\ref{cond2-}) the value $
+\infty$ is also excluded.

On the output space $\Omega $, we consider the probability distribution with
the density
\begin{equation}
p_{H}(\omega )=m^{-1}e^{-\theta H_{c}(\omega )},\quad m=\int_{\Omega
}e^{-\theta H_{c}(\omega )}\mu (d\omega )  \label{pF}
\end{equation}
with respect to the measure $\mu $, then
\begin{equation}
h(p_{\rho })=-h(p_{\rho }\Vert p_{H})+\theta \int_{\Omega }p_{\rho }(\omega
)H_{c}(\omega )\mu (d\omega )+\log m.  \label{hrh}
\end{equation}
This implies (\ref{entfin-}) due to the non-negativity of the relative
entropy $h(p_{\rho }\Vert p_{H})$ and the inequality (\ref{cond2-}).

Consider a finite decomposition $\mathcal{V} = \{V_i\}$ of the output space $
\Omega$ and the measurement channel $\mathcal{M}_{\mathcal{V}}$ described by
POVM $\{M(V_i)\}$, which is embedded into a quantum channel with
finite-dimensional output.

For the capacity of $\mathcal{M}_{\mathcal{V}}$ we have
\begin{equation*}
C(\mathcal{M},H,E)\geq C(\mathcal{M}_{\mathcal{V}},H,E)=\sup_{\pi :~\mathrm{
Tr}\bar{\rho}_{\pi }H\leq E}I(\pi ,\mathcal{M}_{\mathcal{V}}),
\end{equation*}
where in the last equality we take a supremum over all ensembles and use the
fact that the measurement channel $\mathcal{M}_{\mathcal{V}}$ can be
considered as a quantum entanglement-breaking channel \cite{h1}. By taking
supremum over the decompositions $\mathcal{V}$ and using the fact that $
\sup\limits_{\mathcal{V}}I(\pi ,\mathcal{M}_{\mathcal{V}})=I(\pi ,{\mathcal{M
}})$ (theorem 2.2 in \cite{dobr}), we obtain the lower estimate which,
together with the upper estimate (\ref{lek}) gives (\ref{cmfe}). $\Box $

\begin{definition}
Let $G$ be a locally compact group, acting continuously on the transitive $G$
-space $\Omega,$ let also $V:~g \rightarrow V_g$ be a continuous
(projective) unitary representation of the group $G$ in the Hilbert space $
\mathcal{H}$. A POVM $M=\{ M(A), A \in \mathcal{F}\}$ in $\mathcal{H}$ is
called \textit{covariant} under $V$, if
\begin{equation*}
V_g^* M(A)V_g = M(g^{-1}A),~A \in \mathcal{F}.
\end{equation*}
\end{definition}

The following statement holds for the corresponding covariant measurement
channel (an analogue of proposition 1 from \cite{h2}). We assume that the
channel ${\mathcal{M}}$ satisfies the conditions of proposition \ref{p1} so
that the formula (\ref{cmfe}) holds.

\begin{proposition}
\label{p2} \textit{\ Let the following conditions be satisfied for the
measurement channel $\mathcal{M}$ corresponding to the observable $M$ which
is covariant under an irreducible square integrable representation $g
\rightarrow V_g$ of a unimodular group $G$:}

1.$\sup\limits_{\rho:~\mathrm{Tr} \rho H \leq E} h(\mathcal{M}(\rho))$ \textit{
is attained on a state} $\rho_E^0$;

2. $\inf\limits_{\rho} h(\mathcal{M}(\rho))$ \textit{is attained on a state} $
\rho_0$;

3. \textit{there exists a Borel probability measure $\pi_E^0$ on $G$ such that}
\begin{equation*}
\rho_E^0 = \int_G V_g \rho_0 V_g^* \pi_E^0(dg).
\end{equation*}

\textit{Then the capacity of the channel $\mathcal{M}$ is given by the formula
\begin{equation}  \label{ccov}
C(\mathcal{M}, H, E) = h(\mathcal{M}(\rho_E^0)) - h(\mathcal{M}(\rho_0)),
\end{equation}
and it is attained on the ensemble $\{ \pi_E^0(dg), V_g \rho_0 V_g^*\}$.}
\end{proposition}

\textit{Proof.} From (\ref{cmfe})
\begin{equation}
C(\mathcal{M},H,E) \leq h(\mathcal{M}(\rho _{E}^{0}))-h(\mathcal{M}(\rho
_{0})).  \label{ner}
\end{equation}
For a fixed point $\omega _{0}\in \Omega $, consider its stationary subgroup
$G_{0}$. According to theorem 4.8.3 of \cite{asp} the relation
\begin{equation}
M(A)=\int_{A}V_{g^{\prime }}P_{0}V_{g^{\prime }}^{\ast }\,\mu (d\omega ),
\mbox{ where }\omega =g^{\prime }\omega _{0}  \label{2.2}
\end{equation}
establishes one-to-one correspondence between the measurements $\{M(A)\},$
covariant with respect to the irreducible square integrable representation $
g\rightarrow V_{g}$ of the unimodular group $G$, and the density operators $
P_{0},$ commuting with operators $\{V_{g};g\in G_{0}\}$. Here $\mu (d\omega
) $ is properly normalized $G$-invariant measure on $\Omega$, and the integral (\ref{2.2})
is understood in the weak sense.

Using the representation (\ref{2.2}), the covariance of the measurement
channel $\mathcal{M}$ and the invariance of the measure $\mu $, we get
\begin{equation*}
h(\mathcal{M}(V_{g}\rho _{0}V_{g}^{\ast }))=h(\mathcal{M}(\rho _{0})),
\end{equation*}
for any $g\in G$. Substituting the ensemble $\{\pi _{E}^{0}(dg),V_{g}\rho
_{0}V_{g}^{\ast }\}$ in the formula (\ref{cmfe}), we obtain that the upper
bound (\ref{ner}) is achieved, which completes the proof. $\Box $

\section{The classical capacity of general Gaussian observable}

\label{s2}

\bigskip The measurement channels we consider in this section correspond to
general Gaussian observables in the sense of \cite{h} (linear measurements
in \cite{asp}, see also \cite{acc} for the gauge covariant case). In what
follows the quantum input space $\mathcal{H}$ will be the Hilbert space of
an irreducible representation of the canonical commutation relations (\ref
{weyl}) (see Appendix), and the classical output space $\Omega =Z=\mathbb{R}
^{2s}.$

We will consider the general Gaussian measurement channel $\rho \rightarrow
\widetilde{\mathcal{M}}[\rho ]$, described by POVM on $Z$
\begin{equation}
\widetilde{M}(d^{2s}z)=D(Kz)\rho _{\beta }D(Kz)^{\ast }\frac{\left\vert \det
K\right\vert \,d^{2s}z}{\left( 2\pi \right) ^{s}};\quad z\in \mathbb{R}^{2s},
\label{MTA}
\end{equation}
where $z$ is $2s-$dimensional real vector running in the symplectic space $(
\mathbb{R}^{2s},\Delta )$, $D(z)=W(-\Delta ^{-1}z)$ are the displacement
operators (see Appendix) satisfying the equation that follows from the
canonical commutation relations (\ref{weyl})
\begin{equation*}
D(z)^{\ast }W(w)D(z)=\exp \left( iw^{t}z\right) W(w),
\end{equation*}
$K$ is a nondegenerate real matrix and $\rho _{\beta }$ is a centered
Gaussian density operator with the real symmetric covariance matrix $\beta .$
In \cite{acc} it is shown that without loss of generality we can put $K=I$
and consider the POVM
\begin{equation}
M(d^{2s}z)=D(z)\rho _{\beta }D(z)^{\ast }\frac{d^{2s}z}{\left( 2\pi \right)
^{s}}  \label{MTB}
\end{equation}
and the corresponding measurement channel $\mathcal{M}$. In our case $\mu$
is just the normalized Lebesgue measure on $Z=\mathbb{R}^{2s}.$
The fact that the formulas (\ref{MTB}), (\ref{MTA}) determine POVM follows
from theorem 4.8.3 of \cite{asp}.

Consider the quadratic Hamiltonian
\begin{equation}
H=R\epsilon R^{t},  \label{ham}
\end{equation}
where $R$ is the vector of canonical observables defined by (\ref{cano}), $
\epsilon =\left[ \epsilon _{jk}\right] $ is positive definite real symmetric
matrix, so that the mean energy of the input state $\rho $ is equal to
\begin{equation*}
\mathrm{Tr}\rho H=\mathrm{Sp\,}\epsilon \alpha ,
\end{equation*}
where $\mathrm{Sp}$ denotes trace of $2s\times 2s-$matrices as distinct from
the trace of operators in the Hilbert space and
\begin{equation*}
\alpha =\mathrm{Re\,}\mathrm{Tr\,}R^{t}\rho R
\end{equation*}
is the covariance matrix of the state $\rho$. Then the input energy
constraint has the form $\mathrm{Sp\,}\epsilon \alpha \leq E,$ where $E$ is
a positive number. Let us show that in the Gaussian case we are considering,
the conditions of propositions \ref{p1}, \ref{p2} are fulfilled allowing to
compute the \textit{energy-constrained classical capacity} of the channel $
\mathcal{M}$.

\begin{lemma}
\textit{The conditions of proposition \ref{p1} are fulfilled for any positive
definite quadratic form} $H_{c}(z)$.
\end{lemma}

\textit{Proof}. All such forms are equivalent in the sense $H_{c,1}(z)\asymp
H_{c,2}(z)\ $ for all $z,$ i.e. there exist positive constants $k_{1},k_{2},$
such that $k_{1}H_{c,1}(z)\leq H_{c,2}(z)\leq k_{2}H_{c,1}(z).$ Therefore it
is sufficient to prove the inequality
\begin{equation}
\int_{\mathbb{R}^{2s}}H_{c}(z)\,M(d^{2s}z)=\int_{\mathbb{R}
^{2s}}H_{c}(z)\,D(z)\rho _{\beta }D(z)^{\ast }\frac{d^{2s}z}{\left( 2\pi
\right) ^{s}}\leq c_{1}H+c_{2}I  \label{rr}
\end{equation}
where $c_{1}, c_{2}>0,$ and $H$ is given by (\ref{ham}), for at least one
such form $H_{c}(z)$.

Choose a symplectic basis $\left\{ e_{j},h_{j}\right\} $ associated with the
matrix $\beta $ (see Appendix, with $\alpha $ replaced by $\beta $) and let
the real $2s-$vector $z$ be represented by its coordinates in the
decomposition
\begin{equation*}
z=\sum_{j=1}^{s}x_{j}e_{j}+y_{j}h_{j},
\end{equation*}
so that
\begin{equation*}
Rz=\sum_{j=1}^{s}x_{j}\tilde{q}_{j}+y_{j}\tilde{p}_{j},
\end{equation*}
where $\tilde{q}_{j}=Re_{j},\,\tilde{p}_{j}=Rh_{j}.$ Consider the
corresponding complex representation of the symplectic space, where $z$ is
replaced by the complex $s-$vector $\mathbf{z}$ with the components $z_{j}=
\frac{1}{\sqrt{2}}\left( x_{j}+iy_{j}\right) $ (see e.g. \cite{acc}). We
will prove the identity
\begin{equation}  \label{ide}
\int_{\mathbb{C}^{s}}\mathbf{z}^{\ast }\mathbf{z}D(\mathbf{z})\rho _{\beta
}D(\mathbf{z})^{\ast }\frac{d^{2s}\mathbf{z}}{\pi ^{s}}=\tilde{H}+c_{2}I.
\end{equation}
where $\tilde{H}=\frac{1}{2}\sum_{j=1}^{s}\left( \tilde{q}_{j}^{2}+\tilde{p}
_{j}^{2}\right) $ and $c_{2}>0.$ This implies (\ref{rr}) because $\tilde{H}$
is a positive definite quadratic form in $R$, namely $\tilde{H}=RQR^{t},$
with $Q=\frac{1}{2}\sum_{j=1}^{s}\left( e_{j}e_{j}^{t}+h_{j}h_{j}^{t}\right)
\leq c_{1}I_{2s},$ the symbol $I_{2s}$ denotes the unit $2s\times 2s$-matrix
and $c_{1}=\left\Vert Q\right\Vert .$

The density operator $\rho _{\beta }$ admits $P$-representation
\begin{equation*}
\rho _{\beta }=\int_{\mathbb{C}^{s}}\left\vert \mathbf{w}\right\rangle
\left\langle \mathbf{w}\right\vert \exp \left[ -\mathbf{w}^{\ast }N_{\beta
}^{-1}\mathbf{w}\right] \frac{d^{2s}\mathbf{w}}{\pi ^{s}\det N_{\beta }}
\end{equation*}
where $N_{\beta }$ is the diagonal matrix with the entries $N_{j}>0,
\,j=1,\dots ,s,$ (see e.g. \cite{asp}). The quantity in the left hand side
of (\ref{ide}) is the same as
\begin{equation*}
\int_{\mathbb{C}^{s}}\int_{\mathbb{C}^{s}}\mathbf{z}^{\ast }\mathbf{z}
\left\vert \mathbf{z+w}\right\rangle \left\langle \mathbf{z+w}\right\vert
\exp \left[ -\mathbf{w}^{\ast }N_{\beta }^{-1}\mathbf{w}\right] \frac{d^{2s}
\mathbf{w}}{\pi ^{s}\det N_{\beta }}\frac{d^{2s}\mathbf{z}}{\pi ^{s}}.
\end{equation*}
By changing variable $\mathbf{z+w\rightarrow z}$ and taking into account
zero first moments of the Gaussian distribution, we obtain that this is
equal to
\begin{eqnarray*}
&&\int_{\mathbb{C}^{s}}\int_{\mathbb{C}^{s}}\left( \mathbf{z-w}\right)
^{\ast }\left( \mathbf{z-w}\right) \left\vert \mathbf{z}\right\rangle
\left\langle \mathbf{z}\right\vert \exp \left[ -\mathbf{w}^{\ast }N_{\beta
}^{-1}\mathbf{w}\right] \frac{d^{2s}\mathbf{w}}{\pi ^{s}\det N_{\beta }}
\frac{d^{2s}\mathbf{z}}{\pi ^{s}} \\
&=&\int_{\mathbb{C}^{s}}\sum_{j=1}^{s}z_{j}\left\vert \mathbf{z}
\right\rangle \left\langle \mathbf{z}\right\vert \bar{z}_{j}\frac{d^{2s}
\mathbf{z}}{\pi ^{s}}+\int_{\mathbb{C}^{s}}\mathbf{w}^{\ast }\mathbf{w}\exp
\left[ -\mathbf{w}^{\ast }N_{\beta }^{-1}\mathbf{w}\right] \frac{d^{2s}
\mathbf{w}}{\pi ^{s}\det N_{\beta }}I,
\end{eqnarray*}
where we used the identity
\begin{equation*}
\int_{\mathbb{C}^{s}}\left\vert \mathbf{z}\right\rangle \left\langle \mathbf{
z}\right\vert \frac{d^{2s}\mathbf{z}}{\pi ^{s}}=I.
\end{equation*}
Taking into account that $z_{j}\left\vert \mathbf{z}\right\rangle =\tilde{a}
_{j}\left\vert \mathbf{z}\right\rangle ,$ where $\tilde{a}_{j}=\frac{1}{
\sqrt{2}}\left( \tilde{q}_{j}+i\tilde{p}_{j}\right) ,$ and using the second
moments of the Gaussian distribution, we obtain
\begin{eqnarray*}
\int_{\mathbb{C}^{s}}\sum_{j=1}^{s}\tilde{a}_{j}\left\vert \mathbf{z}
\right\rangle \left\langle \mathbf{z}\right\vert \tilde{a}_{j}^{\dagger }
\frac{d^{2s}\mathbf{z}}{\pi ^{s}}+\left( \mathrm{Sp\,}N_{\beta }\right) I
&=&\sum_{j=1}^{s}\tilde{a}_{j}\mathbf{\,}\tilde{a}_{j}^{\dagger }+\left(
\mathrm{Sp\,}N_{\beta }\right) I \\
&=&\frac{1}{2}\sum_{j=1}^{s}\left( \tilde{q}_{j}^{2}+\tilde{p}
_{j}^{2}\right) +\left( s/2+\mathrm{Sp\,}N_{\beta }\right) I.
\end{eqnarray*}
Putting $c_{2}=\left( s/2+\mathrm{Sp\,}N_{\beta }\right) ,$ we obtain (\ref
{ide}). $\square $

Gaussian measurement channel (\ref{MTB}) is covariant with respect to the irreducible
(projective) representation $z\rightarrow D(z)$ of the additive group $G=Z$
in the sense
\begin{equation*}
D(z)^{\ast }M(B)D(z)=M(B-z),\quad z\in Z
\end{equation*}
for $M$ given by (\ref{MTB}), as follows from (\ref{weyl}). Notice that $
G_{0}$ is trivial in this case. We will now show that the channel satisfies
the conditions of the proposition \ref{p2}, in particular n. 3 is fulfilled
under certain \textquotedblleft threshold condition\textquotedblright , in
which case the formula (\ref{ccov}) holds.

\bigskip Probability density of outcomes of the observable (\ref{MTB}) for
the Gaussian input state $\rho _{\alpha }$ is

\begin{eqnarray*}
p_{\rho _{\alpha }}(z) &=&\mathrm{Tr\,}\rho _{\alpha }D(z)\rho _{\beta
}D(z)^{\ast } \\
&=&\int \exp \left( -\frac{1}{2}w^{t}\alpha w\right) \exp \left( -iw^{t}z-
\frac{1}{2}w^{t}\beta w\right) \frac{\,d^{2s}w}{\left( 2\pi \right) ^{s}} \\
&=&\frac{1}{\sqrt{\left( 2\pi \right) ^{s}\mathrm{\det }\left( \alpha +\beta
\right) }}\exp \left( -\frac{1}{2}z^{t}\left( \alpha +\beta \right)
^{-1}z\right) .
\end{eqnarray*}
The Parseval's formula was used here for the Weyl transform (theorem 5.3.3
of \cite{asp}).

Denote by $\mathfrak{S}(\alpha )$ the set of all states which have zero
first moments, and finite second moments with the covariance matrix $\alpha
. $ Take an ensemble $\pi $, such that $\bar{\rho}_{\pi }\in \mathfrak{S}
(\alpha ).$ Then $p_{\bar{\rho}_{\pi }}$ is centered probability density
with the covariance matrix $\alpha +\beta $, and arguing similarly to the
proof of theorem 1 in \cite{acc}, we get that the maximum of $h\left( p_{
\bar{\rho}_{\pi }}\right) $, equal to
\begin{equation}
h\left( p_{\rho _{\alpha }}\right) =\frac{1}{2}\log \mathrm{\det }\left(
\alpha +\beta \right) +C,  \label{hps}
\end{equation}
where the constant $C$ depends on the normalization of the Lebesgue measure
involved in the calculation of the differential entropy (see formula (\ref
{dent}) in Appendix), is attained on the Gaussian state $\bar{\rho}_{\pi
}=\rho _{\alpha }$.

Making additional maximization with respect to Gaussian states with
covariance matrix $\alpha $ satisfying the energy constraint $\mathrm{Sp\,}
\epsilon \alpha \leq E$, we obtain
\begin{equation*}
\max_{\rho :\mathrm{Tr}\rho H\leq E}h\left( p_{\rho }\right) =\max_{\alpha
:\,\mathrm{Sp}\alpha \epsilon \leq E}\frac{1}{2}\log \mathrm{\det }\left(
\alpha +\beta \right) +C,
\end{equation*}
and the maximizer is a centered Gaussian state $\rho _{E}^{0}$ with the
covariance matrix
\begin{equation}
\alpha _{E}^{0}=\arg \max_{\alpha :\,\mathrm{Sp}\alpha \epsilon \leq E}
\mathrm{\det }\left( \alpha +\beta \right) .  \label{aeo}
\end{equation}
Thus, n. 1 of proposition 2 follows.

The statement of n. 2 follows from the results concerning the minimal output
entropy. The result of the paper \cite{ghm} (Proposition 4; see also \cite{acc}) concerning
the minimal output entropy of the Gaussian measurement channel implies that the minimizer
$\rho _{0}$ can be taken as the vacuum state related to the complex
structure $J_{\beta }$ (see Appendix). Substituting $\alpha =\frac{1}{2}
\Delta J_{\beta }$ into (\ref{hps}), we get
\begin{equation*}
\min_{\rho }h\left( p_{\rho }\right) =h\left( p_{\rho _{0}}\right) =\frac{1}{
2}\log \mathrm{\det }\left( \beta +\frac{1}{2}\Delta J_{\beta }\right) +C.
\end{equation*}

The condition n. 3 is fulfilled provided
\begin{equation}
{\alpha }_{E}^{0}\geq \frac{1}{2}\Delta J_{\beta }.  \label{thr}
\end{equation}
Indeed, in this case
\begin{equation*}
\rho _{E}^{0}=\int_{\mathbb{R}^{2s}}D(z)\rho _{0}D(z)^{\ast }\,{\pi }
_{E}^{0}(dz),
\end{equation*}
where ${\pi }_{E}^{0}(dz)$ is the centered Gaussian distribution on $Z$ with
the covariance matrix ${\alpha }_{E}^{0}-\frac{1}{2}\Delta J_{\beta }.$ One
can check this by comparing the quantum characteristic functions of both
sides. The matrix inequality (\ref{thr}) is an analog of the
\textquotedblleft threshold condition\textquotedblright\ for the multi-mode
quantum Gaussian channel \cite{h2} (in the case of
one mode inequalities of this kind appeared in \cite{lupo1}, \cite{lupo2}, \cite{schaefer}).

Thus we have proved the following result extending theorem 1 of \cite{acc}
to the case without global gauge symmetry.

\begin{theorem}
\label{tt2} \textit{Let $\widetilde{\mathcal{M}}$ be the measurement channel
corresponding to the Gaussian POVM (\ref{MTA}). Assume that $\alpha _{E}^{0}$
, given by the formula (\ref{aeo}), and $\beta $ satisfy the condition (\ref
{thr}). Then}
\begin{eqnarray}
C (\widetilde{\mathcal{M}},H,E) &=&\frac{1}{2}\log \mathrm{\det }
\left( \alpha _{E}^{0}+\beta \right) -\frac{1}{2}\log \mathrm{\det }\left(
\beta +\frac{1}{2}\Delta J_{\beta }\right)  \label{CXA} \\
&=&\frac{1}{2}\log \mathrm{\det }\left[ I+\left( \alpha _{E}^{0}-\frac{1}{2}
\Delta J_{\beta }\right) \left( \beta +\frac{1}{2}\Delta J_{\beta }\right)
^{-1}\right] ,  \nonumber
\end{eqnarray}
\textit{which is attained on the ensemble of $J_{\beta }-$coherent states $
D(z)\rho _{0}D(z)^{\ast }$ (see Appendix), where $z$ has the centered
Gaussian probability distribution with the covariance matrix} ${\alpha
_{E}^{0}}-\frac{1}{2}\Delta J_{\beta }.$
\end{theorem}

Finding the covariance matrix (\ref{aeo}) is a separate finite-dimensional optimization problem
which can be solved analytically in some special cases.

\textbf{Example}: Take the energy operator of the signal mode $H=(q^{2}+p^{2})/2$
with the corresponding complex structure $J_{H}$ $=\left[
\begin{array}{cc}
0 & -1 \\
1 & 0
\end{array}
\right] .$ We consider the POVM described by the formula (\ref{MTB}) where
the covariance matrix of the quantum Gaussian noise is
\begin{equation*}
\beta =\left[
\begin{array}{cc}
\beta _{1} & 0 \\
0 & \beta _{2}
\end{array}
\right] ;\quad \beta _{1}\beta _{2}\geq \frac{1}{4}.
\end{equation*}
In quantum optics this would correspond to the heterodyne measurement of the
signal mode with the squeezed quantum noise from the local oscillator.

The complex structure of the measurement noise is
\begin{equation*}
J_{\beta }=\left[
\begin{array}{cc}
0 & -\sqrt{\beta _{2}/\beta _{1}} \\
\sqrt{\beta _{1}/\beta _{2}} & 0
\end{array}
\right] ,
\end{equation*}
which does not commute with $J_{H}$ unless $\beta _{1}=\beta _{2}.$ The
covariance matrix of the squeezed vacuum is (see Appendix)
\begin{equation*}
\frac{1}{2}\Delta J_{\beta }=\frac{1}{2}\left[
\begin{array}{cc}
\sqrt{\beta _{1}/\beta _{2}} & 0 \\
0 & \sqrt{\beta _{2}/\beta _{1}}
\end{array}
\right] .
\end{equation*}
and
\begin{equation*}
\beta +\frac{1}{2}\Delta J_{\beta }=\left[
\begin{array}{cc}
\beta _{1}+\frac{1}{2}\sqrt{\beta _{1}/\beta _{2}} & 0 \\
0 & \beta _{2}+\frac{1}{2}\sqrt{\beta _{2}/\beta _{1}}
\end{array}
\right] ,
\end{equation*}
so that $\det \left( \beta +\frac{1}{2}\Delta J_{\beta }\right) =\left(
\sqrt{\beta _{1}\beta _{2}}+1/2\right) ^{2},$ hence the second term in (\ref
{CXA}) is $-\log \left( \sqrt{\beta _{1}\beta _{2}}+1/2\right) .$

To compute the first term, we can restrict to diagonal input covariance
matrices
\begin{equation*}
\alpha =\left[
\begin{array}{cc}
\alpha _{1} & 0 \\
0 & \alpha _{2}
\end{array}
\right] ,\quad \alpha _{1}+\alpha _{2}\leq 2E,\quad \alpha _{1}\alpha
_{2}\geq \frac{1}{4}.
\end{equation*}
The matrix
\begin{equation*}
\alpha +\beta =\left[
\begin{array}{cc}
\beta _{1}+\alpha _{1} & 0 \\
0 & \beta _{2}+\alpha _{2}
\end{array}
\right]
\end{equation*}
has the determinant $\left( \beta _{1}+\alpha _{1}\right) \left( \beta
_{2}+\alpha _{2}\right),$ so that the maximized expression is $\log \sqrt{
\left( \beta _{1}+\alpha _{1}\right) \left( \beta _{2}+\alpha _{2}\right) }.$
Since $\log x$ is increasing function, we have to maximize $\left( \beta
_{1}+\alpha _{1}\right) \left( \beta _{2}+\alpha _{2}\right) $ under the
constraints $\alpha _{1}+\alpha _{2}\leq 2E,\quad \alpha _{1}\alpha _{2}\geq
\frac{1}{4}.$ The first constraint gives the values
\begin{equation*}
\alpha _{1}^{0}=E+\left( \beta _{2}-\beta _{1}\right) /2,\quad \alpha
_{2}^{0}=E-\left( \beta _{2}-\beta _{1}\right) /2,
\end{equation*}
corresponding to the maximal value of the first term in (\ref{CXA})
\begin{equation*}
\log \frac{1}{2}\left(2E+\left( \beta _{1}+\beta _{2}\right) \right) .
\end{equation*}
The second constraint will be automatically fulfilled provided we impose the
condition (\ref{thr}) which amounts to
\begin{equation*}
\alpha _{1}^{0}\geq \frac{1}{2}\sqrt{\beta _{1}/\beta _{2}},\quad\alpha
_{2}^{0}\geq \frac{1}{2}\sqrt{\beta _{2}/\beta _{1}},
\end{equation*}
or
\begin{equation*}
E\geq \frac{1}{2}\left(\max \left\{ \sqrt{\beta _{1}/\beta _{2}},\sqrt{\beta _{2}/\beta _{1}}
\right\} +\left\vert \beta _{2}-\beta _{1}\right\vert\right) .
\end{equation*}
Under this condition
\begin{eqnarray*}
 C(\mathcal{M};H,E) &=& \log \left(E+\left( \beta _{1}+\beta
_{2}\right)/2 \right) -\log \left( \sqrt{\beta _{1}\beta _{2}}+1/2\right) \\
   &=& \log \left( \frac{2E+\left( \beta _{1}+\beta _{2}\right) }{2\sqrt{\beta
_{1}\beta _{2}}+1}\right) .
\end{eqnarray*}

\ack This work is supported by Russian Science
Foundation under the grant No 19-11-00086. The authors are grateful to M.E.
Shirokov for useful remarks.

\section*{Appendix}

In this section we systematically use notations and some results from the
book \cite{h} where further references are given. Consider a
finite-dimensional symplectic space $(Z,\Delta ) $ with $Z=\mathbb{R}^{2s}$,
\begin{equation}
\Delta =\mathrm{diag}\left[
\begin{array}{cc}
0 & 1 \\
-1 & 0
\end{array}
\right]_{j=1,\dots, s} .  \label{delta}
\end{equation}
Let $\mathcal{H}$ be the space of an irreducible representation $
z\rightarrow W(z);\,z\in Z$ of the canonical commutation relations
\begin{equation}
W(z)W(z^{\prime })=\exp [-\frac{i}{2} z^{t}\Delta z^{\prime
}]\,W(z+z^{\prime }).  \label{weyl}
\end{equation}
Here $W(z)=\exp i\,Rz$ are the unitary Weyl operators, where
\begin{equation}  \label{cano}
Rz=\sum_{j=1}^{s}(x_{j}q_{j}+y_{j}p_{j}),
\end{equation}
$z=[x_{j}, y_{j}]_{j=1,\dots,s}^t$, are the canonical observables of the
quantum system.

Operator $J$ in $(Z,\Delta )$ is called \textit{operator of complex structure
} if
\begin{equation}
J^{2}=-I_{2s},  \label{j2e}
\end{equation}
where $I_{2s}$ is the identity operator in $Z$, and it is $\Delta -$positive
in the sense that
\begin{equation}
\Delta J=-J^{t}\Delta ,\quad \Delta J\geq 0.  \label{comstr}
\end{equation}

A centered Gaussian state $\rho _{\alpha }$ on $\mathfrak{B}(\mathcal{H})$
is determined by its covariance matrix $\alpha =\mathrm{Re\,}\mathrm{Tr\,}
R^{t}\rho R$ which is a real symmetric $2s\times 2s$-matrix satisfying
\begin{equation}
\alpha \geq \pm \frac{i}{2}\Delta .  \label{ur}
\end{equation}
This state is pure if and only if $\alpha =\frac{1}{2}\Delta J.$ It is
called $J-$vacuum and denoted $\rho _{0}.$ The non-centered pure states $
D(z)\rho _{0}D(z)^{\ast }$ are called $J-$coherent states (see sec. 12.3.2
of \cite{h}).

Consider the operator $A=\Delta ^{-1}\alpha $. The operator $A$ is
skew-symmetric in the Euclidean space $(Z,\alpha )$ with the scalar product $
\alpha (z,z^{\prime })=z^{t}\alpha z^{\prime }$. According to a theorem from
linear algebra, there is an orthogonal basis $\left\{ e_{j},h_{j}\right\} $
in $(Z,\alpha )$ and positive numbers $\left\{ \alpha _{j}\right\} $ such
that
\begin{equation*}
Ae_{j}=\alpha _{j}h_{j};\quad Ah_{j}=-\alpha _{j}e_{j}.
\end{equation*}
Eq. (\ref{ur}) implies $N_{j}\equiv \alpha _{j}-1/2\geq 0.$ Choosing the
normalization~ $\alpha (e_{j},e_{j})=\alpha (h_{j},h_{j})=\alpha _{j}$ gives
the symplectic basis in $(Z,\Delta )$ with the required properties.

There is at least one operator of complex structure, commuting with the
operator $A=\Delta ^{-1}\alpha ,$ namely, the orthogonal operator $J_{\alpha
}$ from the polar decomposition
\begin{equation}
A=\left\vert A\right\vert J_{\alpha }=J_{\alpha }\left\vert A\right\vert
\label{kspd}
\end{equation}
in the Euclidean space $(Z,\alpha ).$ The action of $J_{\alpha }$ in the
symplectic basis is given by the formula
\begin{equation*}
J_{\alpha }e_{j}=h_{j},\quad J_{\alpha }h_{j}=-e_{j}.
\end{equation*}

We will need the formula for the differential entropy of a nondegenerate
multidimensional Gaussian probability distribution $p_{\alpha }$ with the
covariance matrix $\alpha :$
\begin{equation}
h(p_{\alpha })=\frac{1}{2}\log \det \alpha +C=\frac{1}{2}\mathrm{Sp}\log
\alpha +C,  \label{dent}
\end{equation}
where the constant $C$ depends on the normalization of the Lebesgue measure
involved in the definition of the differential entropy (cf. \cite{cover}).


\section*{References}
\begin{thebibliography}{99}

\bibitem{cover} Cover T M and Thomas J A 1996 \textit{Elements of Information
Theory}, 2nd edition (New York: John Wiley \& Sons)

\bibitem{da} Dall'Arno M, D'Ariano G M and Sacchi M F 2011 Informational power
of quantum measurements \textit{Phys. Rev.} \textbf{A 83}, 062304

\bibitem{dobr} Dobrushin R L 1959 General formulation of Shannon theorem in
information theory \textit{Russian Math. Surveys} \textbf{14}:6 3-104

\bibitem{ghm} Giovannetti V, Holevo A S and Mari A 2015 Majorization and
additivity for multimode bosonic Gaussian channels \textit{Theor. Math. Phys.}
\textbf{182}:2 284-293

\bibitem{hall} Hall M J W 1997 Quantum information and correlation bounds,
\textit{Phys. Rev. A} \textbf{55}:1 1050-2947

\bibitem{ClCap} Holevo A S 2004 Entanglement-assisted capacities of
constrained quantum channels \textit{Theory Probab. Appl.} \textbf{48}:2 243-255

\bibitem{h1} Holevo A S 2008 Entanglement-breaking channels in infinite
dimensions \textit{Problems Inform. Transmission} \textbf{44}:3 171-184

\bibitem{asp} Holevo A S 2011 \textit{Probabilistic and statistical aspects of quantum
theory} 2nd English edition (Pisa:Edizioni Della Normale)

\bibitem{h5} Holevo A S 2012 Information capacity of quantum observable
\textit{Probl. Inform. Transmission} \textbf{48}:1 1-10

\bibitem{h2} Holevo A S 2016 On the constrained classical capacity of
infinite-dimensional covariant channels \textit{J. Math. Phys.} \textbf{57}:1
15203

\bibitem{h} Holevo A S 2019 \textit{Quantum systems channels information: a
mathematical introduction} 2nd edition (Berlin/Boston: De Gruyter)

\bibitem{acc} Holevo A S 2019 Gaussian maximizers for quantum Gaussian
observables and ensembles arXiv:1908.03038 [math-ph]

\bibitem{ContEns} Holevo A S and Shirokov M E 2005 Continuous ensembles and the
capacity of infinite-dimensional quantum channels \textit{Theory Probab. Appl.}
\textbf{50}:1 86-98

\bibitem{Hybrid-2} Kuznetsova A A and Holevo A S 2015 Coding theorems for hybrid
channels. II \textit{Theory Probab. Appl.} \textbf{59}:1 145-154

\bibitem{lupo1} Lupo C, Pilyavets O V and Mancini S 2009 Capacities of lossy bosonic channel with correlated noise \textit{New J. Phys.} \textbf{11} 063023

\bibitem{lupo2} Lupo C, Pirandola S, Aniello P and Mancini S 2011 On the classical capacity of quantum Gaussian channels \textit{Phys. Scr.} \textbf{T 143} 014016

\bibitem{Oreshkov} Oreshkov O, Calsamiglia J, Munoz-Tapia R and Bagan E 2011 Optimal signal states for quantum detectors
\textit{New J. Phys.} \textbf{13} 073032

\bibitem{schaefer} Sch\"{a}fer J, Karpov E, Garc{\'i}a-Patr\'on R,
Pilyavets O V and Cerf N J 2013 Equivalence Relations for the Classical Capacity
of Single-Mode Gaussian Quantum Channels \textit{Phys. Rev. Lett.} \textbf{111}
030503

\end{thebibliography}
\end{document}